\documentclass[pra,twocolumn,amsfonts,amssymb,longbibliography]{revtex4-1}

\usepackage{graphicx}
\usepackage{bm}

\begin{document}

\title{Ground-state pressure of quasi-2D Fermi and Bose gases}
\author{Vasiliy Makhalov}\author{Kirill Martiyanov}\author{Andrey Turlapov}\email{turlapov@appl.sci-nnov.ru}\affiliation{Institute of Applied Physics, Russian Academy of Sciences, ul.~Ulyanova 46, Nizhniy Novgorod, Russia}\date{\today}

\begin{abstract} Using an ultracold gas of atoms, we have realized a quasi-two-dimensional Fermi system with widely tunable \textit{s}-wave interactions nearly in a ground state. Pressure and density are measured. The experiment covers physically different regimes: weakly and strongly attractive Fermi gases and a Bose gas of tightly bound pairs of fermions. In the Fermi regime of weak interactions, the pressure is systematically above a Fermi-liquid-theory prediction, maybe due to mesoscopic effects. In the opposite Bose regime, the pressure agrees with a bosonic mean-field scaling in a range beyond simplest expectations. In the strongly interacting regime, measurements disagree with a purely 2D model. Reported data may serve for sensitive testing of theoretical methods applicable across different quantum physics disciplines.
\end{abstract}

\maketitle

Two-dimensional many-body quantum systems show interesting physics and are technologically important. In 2D the phenomena of superfluidity and Bose condensation become clearly separated~\cite{BerezinskiiIIEng}. High-temperature superconductivity is attributed to the 2D structure of the materials~\cite{LoktevReview2001}. Semiconductor and oxide interfaces containing 2D electron gas are important for modern and prospective electronics~\cite{Review2DEG1982,Oxide2DEG2008}.

The concept of a Bardeen-Cooper-Schrieffer (BCS) to Bose-Einstein-condensate (BEC) crossover~\cite{PitaevskiiFermiReview,RanderiaReview2013} gives a unified view at some Fermi and Bose systems: By varying interactions, a gas of fermions obeying the BCS or similar model may be smoothly converted into a gas of pointlike bosons, which are pairs of the initial fermions. Such a crossover has been predicted for excitons~\cite{ExcitonCrossover1968Eng} and quarks~\cite{QuarkCrossover2002} and realized in a 3D gas of ultracold fermionic atoms with \textit{s}-wave interactions~\cite{Grimmbeta}. Measurements on this system have stimulated development of the many-body quantum theory~\cite{PitaevskiiFermiReview,RanderiaReview2013}, especially for the challenging regime of strong interactions which lies between the BCS and Bose asymptotes.

The 2D BCS-BEC crossover for fermions with \textit{s}-wave interactions is the focus of this Letter. The strongly interacting regime of this crossover may be relevant to high-temperature superconductors: While the superconducting phase of the cuprates has \textit{d}-wave symmetry~\cite{LoktevReview2001}, the \textit{s}-wave symmetry has been detected in the pseudogap phase~\cite{sWavePseudogap2012}. Exploring the Bose part of the crossover compliments studies of interacting 2D Bose gases~\cite{ChengChin2DBose2013} by reaching stronger interactions. Studying the fermionic side may add to the understanding of 2D Fermi liquids. Failure of the mean-field description is an example of theoretical challenges in 2D: In 3D the BCS-BEC crossover is qualitatively modeled by a mean field of Cooper pairs (Fig.~5 of Ref.~\cite{PitaevskiiFermiReview}), while in 2D a similar model is qualitatively incorrect, predicting an interaction-independent equation of state at zero temperature~\cite{Giorgini2D2011}.

The pure 2D paradigm assumes motion strictly in the $xy$ plane and no $z$ dependence in interactions. In reality, particles experience zero-point oscillations along $z$ and interact via 3D potentials. The term ``quasi-2D'' generally indicates some departure from the pure 2D approximation while the kinematics remains close to 2D. For example, in most cuprate superconductors the 2D picture is altered by interlayer hopping of electrons, but pure 2D models are widespread~\cite{LoktevReview2001}. An example of a 2D model insufficiency is $^3$He on a substrate, where an increase of zero-point oscillations relative to the atom-atom interaction range brings about the formation of a self-bound liquid~\cite{Quasi2D3He2013}.

Ultracold Fermi atoms~\cite{BlochLowDReview2008,PitaevskiiFermiReview} are well suited for studying the crossover and testing the applicability of purely 2D models in quasi-2D (Q2D). The atomic system allows an \textit{ab initio} description because of purity and the knowledge of microscopic and external parameters; e.~g., 2D kinematics is achieved by holding atoms in the lowest state of the precisely known potential $m\omega_z^2z^2/2$~\cite{Fermi2D}, where $m$ is the atom mass. The range of atom-atom interaction is nearly zero, which has two consequences: (i)~The \textit{s}-wave collisions are quasi-2D rather than 2D, because at distances $\ll l_z\equiv\sqrt{\hbar/2m\omega_z}$ the two-atom wave function is determined by the 3D scattering length~$a$; (ii)~the interaction may be mapped onto \textit{s}-wave scattering by a purely 2D potential~\cite{Shlyapnikov2DScattering2001}; i.~e., purely 2D collisions are simulated. There is a controversy, however, in calculating $a_2$, the corresponding 2D \textit{s}-wave scattering length, which we resolve below.


We find $a_2$ by equating the amplitude of 2D scattering
\begin{equation}
f_{\text{2D}}(q,a_2)=-\frac{2\pi}{\ln(qa_2e^{\gamma_E}/2i)}
\label{eq:f2}
\end{equation}
to the scattering amplitude of atoms interacting via 3D contact potential and confined to the lowest state of potential $m\omega_z^2z^2/2$~\cite{Shlyapnikov2DScattering2001}:
\begin{equation}
f_{\text{Q2D}}(q,a,l_z)=\frac{2\pi}{\sqrt\pi l_z/a+w(q^2l_z^2)/2}.
\label{eq:f2ho}
\end{equation}
Here $\gamma_E\simeq0.577$ is Euler's constant, $w(\xi)$ is defined in~\cite{Shlyapnikov2DScattering2001} and in Eq.~(\ref{eq:wOfxi}) of Appendix~\ref{sec:DataReps},
and $\hbar q=\sqrt{\mathstrut2\mu m}$ is the relative momentum expressed via the chemical potential $\mu$.

In the alternative approach, similar to Refs.~\cite{Kohl2DPseudogap2011,Kohl2DFermiLiquid2012} and not adopted here, $a_2$ is found from binding energy $E_{\text{3D bound}}$ of the 3D dimer molecule in potential $(2m)\omega_z^2z^2/2$ [Eq.~(\ref{eq:Eb}) of Appendix~\ref{sec:DataReps}],
by equating $E_{\text{3D bound}}$ to the binding energy in a 2D potential: $E_{\text{3D bound}}=-4\hbar^2/me^{2\gamma_E}a_2^2$.

In the limit $\mu\ll\hbar\omega_z$, these two approaches give the same $a_2$ for small $a<0$. The controversy is for small $a>0$, which is seen by considering the mean field of a uniform 2D gas of atoms. The leading-order term is $-2\pi\hbar^2n_2/[m\ln(a_2\sqrt{n_2})]$~\cite{HardCore2DBosons1971}. Plugging in $a_2$ derived from $f_{\text{2D}}=f_{\text{Q2D}}$, one obtains the mean-field value $\sqrt{4\pi}\hbar^2n_2a/(ml_z)$ in agreement with Ref.~\cite{Pricoupenko2DMFvsVariational2004},
while the bound-state-based method yields a much larger $a_2$ overestimating the mean field. This motivates the choice of the amplitude-based approach.

While two-body atom-atom collisions are exactly mapped onto purely 2D interactions, the effect of many-body interactions on dimensionality is unclear. Potentially, strong many-body interactions may alter 2D kinematics by populating excited states of motion along $z$, making the system quasi-2D.

The state of tunable atomic Fermi gases with predominantly 2D kinematics has been studied by means of radio-frequency spectroscopy~\cite{Kohl2DPseudogap2011,Zwierlein2DPairing2012,ThomasRFQuasi2D2012,Kohl2DFermiLiquid2012}, measurement of cloud size~\cite{ValeCloudSize2011}, and observing collective modes~\cite{Kohl2DBreathingMode2012}. Experiments~\cite{Kohl2DPseudogap2011,Zwierlein2DPairing2012} have shown that reduced dimensionality makes pairing more favorable. In Ref.~\cite{Kohl2DPseudogap2011}, pair-breaking energy in a strongly interacting Fermi system is in agreement with the mean field of Cooper pairs. Alternatively, in Ref.~\cite{ThomasRFQuasi2D2012}, the excitations are inconsistent with the mean-field interpretation and the system is described as a gas of noninteracting polarons. A quantitative interpretation of these and other finite-temperature studies in the Bose and strongly interacting regimes is complicated, because in 2D quantitative thermometry has been available only for weakly interacting Fermi gas~\cite{Fermi2D,Kohl2DPseudogap2011}. Observation of many-body effects by means of rf spectroscopy puts stringent requirements on experimental precision, because many-body physics is masked by one- and two-body effects.

In this Letter, we report on the controllable realization and study of the quasi-2D BCS-BEC crossover. The thermometry limitations are circumvented by preparing the system nearly in the ground state. The 2D pressure per spin state $P_2$ and the respective numerical planar density $n_2$ are measured. Local thermodynamic quantities have been measured in the 3D BCS-BEC crossover~\cite{JosephSound,SalomonEOS}. In 2D, unlike in 3D, such quantities are sensitive to beyond-mean-field effects even at the qualitative level. In particular, as the system becomes more bosonic, the pressure should drop, contrary to the mean-field expectations~\cite{Giorgini2D2011}.

The apparatus and gas preparation are generally described in Refs.~\cite{FermionsInLattice,Fermi2D} with relevant details elaborated in Appendix~\ref{sec:ExpSetupPrep}.
Lithium-6 atoms are equally populating two lowest-energy spin states $|1\rangle$ and $|2\rangle$. The \textit{s}-wave interactions are controlled by external magnetic field $B$, by using a broad Fano-Feshbach resonance, which in 3D lies at $B=832$~G \cite{JochimNewLiFeshbach2013}. The pancake-shaped trapping potential $V$ is nearly harmonic: $V(\vec{\rho},z)\simeq(\omega_x^2x^2+\omega_y^2y^2+\omega_z^2z^2)m/2$, and tight along $z$: $\omega_z/\omega_\perp=52.2\gg1$, where $\omega_\perp\equiv\sqrt{\mathstrut\omega_x\omega_y}$ and $\omega_y/\omega_x=1.50$. A series of such nearly identical potentials is formed by antinodes of a standing wave; 100--200 adjacent traps are loaded. The longitudinal frequency is chosen in the range $\omega_z/2\pi=2.28$--$13.7$ kHz corresponding to the lattice depth $V_0=(1.9$--$11.6)\hbar\omega_z$. The number of atoms per spin state $N$ varied between 180 and 1040. This gives the noninteracting-gas Fermi energy $E_F=\hbar\omega_\perp\sqrt{2N}=(0.36$--$0.87)\hbar\omega_z$.

\begin{figure}[htb!]
\begin{center}
\includegraphics[width=3.3in]{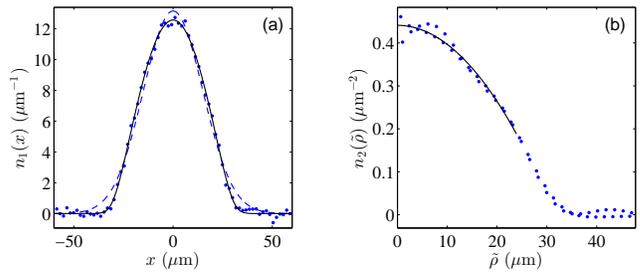}
\end{center}
\caption{(a) Linear density profile $n_1(x)$, no noise filtering applied. Dots are the data for $a_2\sqrt{n_2}=0.89$ ($B=850$ G, $N=500\pm10$, $l_z=-0.29\,a$). Solid and dashed curves are the fits by Thomas-Fermi and Gaussian distribution, respectively. The Gaussian fit is off, which proves deep degeneracy.
(b) Surface density $n_2(\tilde{\rho})$ derived from noise-filtered $n_1(x)$. Dots are the data. The curve is the fit of the parabola $n_2(\tilde{\rho})=n_2-\tilde{\rho}^2n_2''/2$ to the data, which yields the central density $n_2\equiv n_2(\tilde{\rho}=0)$.}
\label{fig:DensityProfiles}
\end{figure}

The pressure and density are measured in the locally homogeneous part of the cloud, near the center $\rho=0$, by analyzing the linear density profiles $n_1(x)$ as the one in Fig.~\ref{fig:DensityProfiles}(a).
%
These profiles are obtained by imaging
the flat clouds from a side, along the $y$ direction, which integrates the density giving the linear distribution $n_1(x)=\int n_2(\vec{\rho})\,dy$ (Appendix~\ref{sec:ImagingDensityProfiles}). We average $n_1(x)$ over 15--30 nearly identical clouds.

The local pressure is obtained from the force balance equation
$\nabla_\perp P_2(\vec{\rho})=-n_2(\vec{\rho})\nabla_\perp V(\vec{\rho},z)$.
Integrating, one finds the central pressure $P_2=m\omega_\perp^2N\left(1-m\omega_x^2\langle x^2\rangle/V_0\right)/2\pi$,
where the transverse potential is expanded up to the quartic term and $\langle x^2\rangle=\frac1N\int x^2n_1(x)dx$.
The planar density profile $n_2(\vec\rho)$ is found by performing, first, noise filtering of $n_1(x)$ and then the inverse Abel transform adjusted for elliptic clouds (Appendix~\ref{sec:ImagingDensityProfiles}).
In Fig.~\ref{fig:DensityProfiles}(b) one may see the planar density distribution in stretched coordinates $\vec{\tilde\rho}=(x,y\,\omega_y/\omega_x)$, in which the clouds are cylindrically symmetric. For Eq.~(\ref{eq:f2ho}), the value of $\mu$ is needed. The scale-invariant assumption $\mu\propto n_2$ together with $dP_2=n_2\,d\mu$ gives the estimate $\mu=E_F\sqrt{P_2/P_{2\,\text{ideal}}}$, where $P_{2\,\text{ideal}}=\pi n_2^2\hbar^2/m$ is the pressure of an ideal Fermi gas of the same density $n_2$. This estimate is of sufficient precision, because departure from $\mu\propto n_2$ is small (Appendix~\ref{sec:ErrAnalysis})
and the function $w(\xi)$ is slow.

\begin{figure}[htb!]
\begin{center}
\includegraphics[width=3.2in]{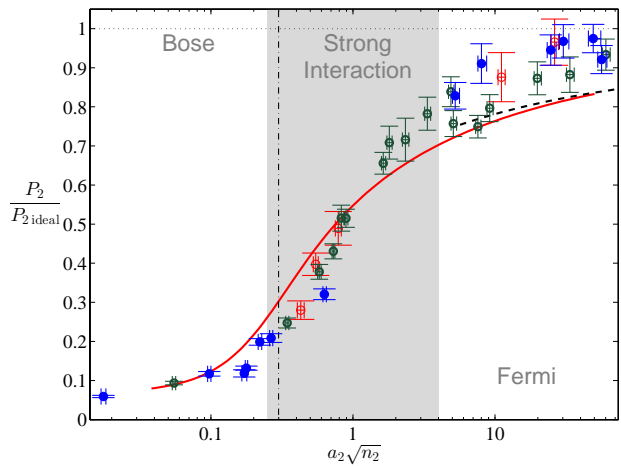}
\end{center}
\caption{Normalized local pressure vs interaction parameter. Data are coded in color by the atom number $N$: smallest $N=220$--$350$ [$E_F=(0.40$--$0.51)\hbar\omega_z$] are shown by empty red circles, intermediate $N=410$--$560$ [$E_F=(0.54$--$0.64)\hbar\omega_z$] by green circles with a thicker border, and largest $N=640$--$910$ [$E_F=(0.69$--$0.82)\hbar\omega_z$] by blue solid dots. Solid red curve: Smooth approximation of the pure 2D Monte Carlo simulation~\cite{Giorgini2D2011}. Dashed curve: Model~\cite{Bloom1975} of a homogeneous 2D Fermi liquid at $T=0$. Dotted line: Mean-field model based on the BCS state~\cite{Randeria2DCrossover1989prl}. The vertical dash-dotted line separates data into two regions: $a_2<l_z$ on the left and $a_2>l_z$ on the right. The data in table form are at the end of the Appendices.
}
\label{fig:PvskFa2}
\end{figure}

The dependance of the normalized pressure on the interaction parameter $a_2\sqrt{n_2}$ is shown in Fig.~\ref{fig:PvskFa2}, which is the main result of the Letter.
The error bars include statistical and systematic errors (Appendix~\ref{sec:ErrAnalysis}).
All known systematic effects, except finite temperature, are corrected for as explained in
Appendix~\ref{sec:ErrAnalysis}.

Interaction parameter $a_2\sqrt{n_2}$ is the ratio of the scattering spatial scale to the interparticle distance. Three regimes may be distinguished: (i) For $a_2\sqrt{n_2}\gg1$, the system is fermionic because the Fermi pressure dominates over the interactions; (ii) at $a_2\sqrt{n_2}\sim1$, the interaction energy is comparable to the Fermi energy, and the system is strongly interacting; (iii) for $a_2\sqrt{n_2}\ll1$, the pairing energy is even larger, and the fermions are bound into compact bosonic pairs; the interaction of pairs is small in comparison to the Fermi energy; therefore, the system is a weakly repulsive Bose gas of molecules. The borders between regimes may be taken approximately at $a_2\sqrt{n_2}=1/4$ and 4 (Fig.~\ref{fig:PvskFa2}). In further discussion, the borders are attached to the data points, which are closest to these two values.



In the Fermi region $a_2\sqrt{n_2}\geqslant4.9$, the ratio $P_2/P_{2\,\text{ideal}}$ is approaching unity as expected (Fig.~\ref{fig:PvskFa2}). Temperatures are evenly distributed in the range $T=(0.02$--$0.15)E_F$, with a confidence interval $\simeq\pm0.03\,E_F$ in each measurement. The temperature in the units of the local Fermi energy $\varepsilon_F=2\pi n_2\hbar^2/m$ is also known, because $\varepsilon_F\simeq E_F$ in this regime. At $T=0.08\,\varepsilon_F$ for a weakly attractive Fermi gas, the gap should be closed for $a_2\sqrt{n_2}\geqslant0.54$~\cite{Shlyapnikov2DFermi2003}. Thus, in the Fermi region the system is likely a Fermi liquid.

For $a_2\sqrt{n_2}\geqslant4.9$, the pressure on average is 10\% above the prediction for a homogeneous 2D Fermi liquid~\cite{Bloom1975}, however.
Neither an unaccounted population of excited states for the motion along $z$ nor a pairing gap would explain higher pressure, because these effects may only reduce $P_2/P_{2\,\text{ideal}}$.
Finite temperature cannot be the reason either: For an ideal uniform Fermi gas, the pressure rises by 2\% when $T/\varepsilon_F$ increases from 0 to 0.08, which gives an estimate for the effect of temperature on the Fermi-liquid pressure. The observed high pressure could be attributed to the mesoscopic character of the system at large $a_2$ values. Whenever $a_2$ is larger than the rms cloud size $\sqrt{\langle\rho^2\rangle}$, the interaction is effectively suppressed, which tunes the gas closer to noninteracting. In the Fermi regime, at the cloud center $a_2/\sqrt{\langle\rho^2\rangle}\simeq a_2\sqrt{n_2}\sqrt{1.5\pi/N}$. For $N=500$, $a_2\sqrt{n_2}=10$ is the crossover point between the locally homogeneous and mesoscopic regimes. According to this criterion, in Ref.~\cite{Kohl2DFermiLiquid2012} the system is mesoscopic for the weakest interactions, but one-particle excitations are found to agree with a model of a finite-temperature locally homogeneous Fermi liquid, which contradicts our pressure measurements.

\begin{figure}[htb!]
\begin{center}
\includegraphics[width=3.2in]{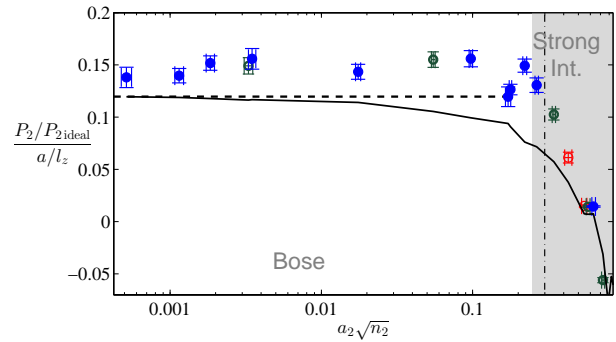}
\end{center}
\caption{Demonstration of linear scaling $P_2/P_{2\,\text{ideal}}\propto a/l_z$ in the Bose regime. Markers: The data $(P_2/P_{2\,\text{ideal}})/(a/l_z)$ vs the interaction parameter. Color coding and the vertical dash-dotted line are the same as in Fig.~\ref{fig:PvskFa2}. Dashed horizontal line: Model for pointlike molecular bosons with 3D interactions, $P_2=P_{2\,\text{ideal}}\frac{0.6a}{l_z\sqrt{8\pi}}$. The solid line connects points calculated via $P_2=P_{2\,\text{Bose}}/2$.}
\label{fig:BoseLinear}
\end{figure}

As the system becomes more bosonic, the ratio $P_2/P_{2\,\text{ideal}}$ decreases following qualitative expectations (Fig.~\ref{fig:PvskFa2}). In the Bose regime $a_2\sqrt{n_2}\ll1$, we find the scaling $P_2/P_{2\,\text{ideal}}\propto a/l_z$ as seen in Fig.~\ref{fig:BoseLinear}, where additional data are also shown in the deep Bose regime $a_2\sqrt{n_2}<0.01$.
To understand this scaling, one may note that the leading term in the pressure of Bose molecules should be the same as for pointlike bosons: $P_{2\,\text{Bose}}=-P_{2\,\text{ideal}}/\ln(a_{2\,\text{mol}}^2n_2)$~\cite{HardCore2DBosons1971}. Here $a_{2\,\text{mol}}$ is the 2D scattering length for molecule-molecule collisions, which may be related to the respective 3D scattering length $a_{\text{mol}}=0.6\,a$~\cite{Petrov} by equating the scattering amplitudes $f_{\text{2D}}(2q,a_{2\,\text{mol}})=f_{\text{Q2D}}(2q,a_{\text{mol}},l_z/\sqrt2)$. The pressure calculated for each datum as $P_2=P_{2\,\text{Bose}}/2$ is shown in Fig.~\ref{fig:BoseLinear} as the broken solid line. To further simplify $P_{2\,\text{Bose}}$, one may take the low-energy limit $2\mu\ll\hbar\omega_z$, yielding $a_{2\,\text{mol}}\simeq2.09\,l_z\exp(-\sqrt{\frac\pi2}\frac{l_z}{a_{\text{mol}}})$, and the limit of unmodified 3D interactions $a_{\text{mol}}\ll l_z$, which all together give $P_{2\,\text{Bose}}\simeq2P_{2\,\text{ideal}}\frac{a_{\text{mol}}}{l_z\sqrt{8\pi}}$ shown by the dashed line in Fig.~\ref{fig:BoseLinear}. This expression gives the scaling $P_2/P_{2\,\text{ideal}}\propto a/l_z$, though in the data the scaling coefficient is $19\%$ higher. Unexpectedly, the data agree with this bosonic mean-field scaling beyond $a_{\text{mol}}\ll l_z$, up to $a_{\text{mol}}/l_z=0.96$ ($a_2\sqrt{n_2}=0.27$) or, in the language of the bosonic coupling parameter $g=-2\pi/\ln(a_{2\,\text{mol}}\sqrt{n_2})$~\cite{ChengChin2DBose2013}, up to $g=2.9$. The agreement extends into the border with the strongly interacting region, where the picture of pointlike bosons is questionable.

In the Bose regime $a_2\sqrt{n_2}\leqslant0.22$, to assure the closeness to the ground state, we measure the temperature fitting the $n_1(x)$ data by bimodal distribution, which is the sum of a Gaussian and zero-temperature Thomas-Fermi distribution $\frac{8N_0}{3\pi R_{\text{TF}}}(1-x^2/R_{\text{TF}}^2)^{3/2}$, where $R_{\text{TF}}$ and $N_0$ are varied. The temperature is inferred from the noninteracting Bose gas relation $N_0/N=1-(T/T_{\text{cr}})^2$, where $T_{\text{cr}}=E_F\sqrt3/\pi$. This procedure may overestimate the temperature~\cite{DalibardBECandBKT2008}. Fitting consistently yields $T<0.5\,T_{\text{cr}}$. To find $T$ in the local units of $\varepsilon_F$, we note that $E_F/\varepsilon_F=\sqrt{P_2/P_{2\,\text{ideal}}}$. Combining this with the asymptote $P_2/P_{2\,\text{ideal}}=0.14\,a/l_z$ seen in Fig.~\ref{fig:BoseLinear}, we obtain the upper bound  $T/\varepsilon_F<0.1\sqrt{a/l_z}$. For the absolute majority of the bosonic data, the temperature is below the Berezinskii-Kosterlitz-Thouless (BKT) transition calculated for weakly interacting bosons~\cite{SvistunovBKT2001}. In a few cases, which we checked in the deep Bose regime, BECs are observed: After release of the gas from the lattice and free expansion for a few milliseconds, straight interference fringes are clearly visible. Indeed, finite Bose systems are predicted to condense~\cite{HardCore2DBosons1971,Kagan2DBEC1987}. The interference, however, has not been studied systematically for all $a_2\sqrt{n_2}$. Since $T/\varepsilon_F$ is low, the actual phase state, whether BKT or BEC, may not affect the pressure significantly: The pressure is continuous over these transitions and, therefore, close to the ground-state value.

In the strongly interacting regime $0.27\leqslant a_2\sqrt{n_2}\leqslant3.3$, quantitative thermometry is presently unavailable. For demonstrating closeness to the ground state, we act along the lines of empirical thermometry~\cite{JointScience}. The density profiles resemble those of the ideal Fermi gas: For the lowest temperatures the edges are sharp, while as $T$ increases the shape transforms into a Gaussian. By fitting the ideal Fermi gas profile~\cite{Fermi2D} to the data $n_1(x)$, as in Fig.~\ref{fig:DensityProfiles}(a), we find the empirical temperature parameter evenly distributed in the range $(T/E_F)_{\text{fit}}=0.02$--$0.16$ indicating deep degeneracy.

For the regime of strong interactions $a_2\sqrt{n_2}\sim1$, in Fig.~\ref{fig:PvskFa2}, one may compare the pressure to the prediction of zero-temperature Monte Carlo model~\cite{Giorgini2D2011} for a uniform gas with 2D atom-atom interactions. Some data are lying on the Monte Carlo curve. When the overall trend is considered, one may see that the slope of the data is steeper (also in Fig.~\ref{fig:PvskFa2Fit} of Appendix~\ref{sec:FittingByCurve}).
Contrary to our findings, measurements of the cloud size~\cite{ValeCloudSize2011} in the strongly interacting regime are reported to quantitatively agree with the Monte Carlo results~\cite{Giorgini2D2011}. If the data of Fig.~\ref{fig:PvskFa2} are plotted by using the definition of $a_2$ adopted in Ref.~\cite{ValeCloudSize2011}, the pressure lies systematically below the Monte Carlo curve in most of the strong-interaction region (Fig.~\ref{fig:PvskFa2Alternatives} of Appendix~\ref{sec:DataReps}).
In Ref.~\cite{ValeCloudSize2011}, the system might be significantly away from the ground state: Unlike here in Fig.~\ref{fig:DensityProfiles}(a), images of the trapped cloud in Ref.~\cite{ValeCloudSize2011} do not show the sharp edge.

\begin{figure}[htb!]
\begin{center}
\includegraphics[width=3.2in]{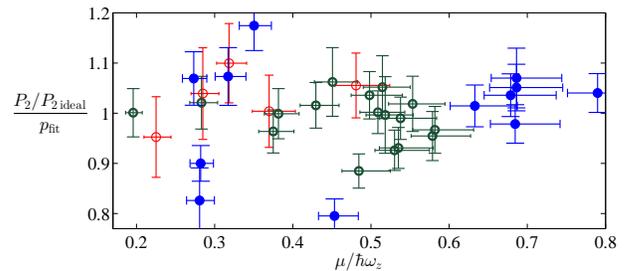}
\end{center}
\caption{Ratio of the measured normalized pressure to the respective values of a smooth fitting function (Appendix~\ref{sec:FittingByCurve})
vs the normalized chemical potential. Color coding is the same as in Fig.~\ref{fig:PvskFa2}.
}
\label{fig:muDoesntMatter}
\end{figure}

In the strongly interacting and Fermi regimes of our experiment, the chemical potential $\mu$ is less than but comparable to $\hbar\omega_z$.
To see whether the closeness of $\mu$ to the excited state matters, we have done measurements with different atom numbers as indicated by color coding in Figs.~\ref{fig:PvskFa2} and \ref{fig:BoseLinear}. Within current precision, there is no dependence on $N$ or $E_F/\hbar\omega_z$. In addition, the data of Fig.~\ref{fig:PvskFa2} are fitted by a smooth curve $p_{2\,\text{fit}}(a_2\sqrt{n_2})$ in the range $0.055\leqslant a_2\sqrt{n_2}\leqslant60$. In Fig.~\ref{fig:muDoesntMatter}, we show the ratio of the measured values to this fit: There are no systematic shifts between the points with low and high $\mu/\hbar\omega_z$.
Also, we rule out two possible sources for the excited state population: (i) There are no thermal excitations~\cite{Fermi2D}; (ii) for two-fermion collisions, scattering into the upper states of motion along $z$ is prohibited by energy and parity conservation, because the collision kinetic energy $2\mu$ is $<2\hbar\omega_z$.
By ruling out the simplest reasons of the excited state population, we do not exclude such a population completely, because it may be induced by strong interactions.
Such quasi-2D effects are a potential reason for deviation from the pure 2D Monte Carlo model~\cite{Giorgini2D2011}.

In conclusion, a widely tunable quasi-2D Fermi system nearly in the ground state has been
realized experimentally. The pressure measurements may be used for sensitive testing of many-body theories including the question of applicability of purely 2D models in strongly interacting systems.

\begin{acknowledgments}
We are thankful to D. S. Petrov for discussions. We acknowledge the financial support by the programs of the Presidium of Russian Academy of Sciences ``Quantum mesoscopic and disordered structures'' and ``Nonlinear dynamics'' and Russian Foundation for Basic Research (Grants No. 11-02-01324-a, No. 11-02-12282-ofi-m-2011, and No. 12-02-31804 mol\_a).
\end{acknowledgments}

\appendix
\section{Ultracold gas in a series of pancake-shaped potentials}\label{sec:ExpSetupPrep}

The gas of lithium-6 atoms is prepared in the potential~\cite{Fermi2D}
\begin{equation}
V(\vec{\rho},z)=sE_{\text{rec}}\left(1-\exp\!\left[-\frac{m(\omega_x^2x^2+\omega_y^2y^2)}{2sE_{\text{rec}}}\right]\,\cos^2kz\right),
\label{eq:OptLattice}
\end{equation}
where $E_{\text{rec}}=\hbar^2k^2/2m$ is the recoil energy [$k=2\pi/(10.6\text{ $\mu$m})$] and $s\equiv V_0/E_{\text{rec}}$ is the dimensionless lattice depth. This potential is due to the dipole force from a standing wave formed by two counterpropagating Gaussian beams with fully overlapping foci, identical power and polarization, and the wavelength of $10.6$ $\mu$m. In Eq.~(\ref{eq:OptLattice}), we neglect the beam divergence, because the Rayleigh length $\simeq3.1$ mm is much bigger than the 200 $\mu$m long region used for the pressure and density measurements. At the bottom of each well, the potential is close to the harmonic shape with $\omega_z=2\sqrt sE_{\text{rec}}/\hbar$.

Preparation of the degenerate gases is similar to that described in Refs.~\cite{FermionsInLattice,Fermi2D}. During the first 8 s of preparation, $10^8$--$10^9$ atoms are collected in a magneto-optical trap (MOT) from an atomic beam. After the MOT fields are turned off, $\sim 10^6$ atoms remain trapped in the standing-wave dipole trap, which spatially overlaps with the MOT and whose depth is $\simeq 280$ $\mu$K. Immediately after the MOT is off, a uniform magnetic field along the $-y$ direction starts to rise reaching 840 G in 100 ms. At 500--600 ms after the MOT is off, one of the beams, forming the standing wave, is gradually extinguished. The gas ends up in a cigar-shaped optical dipole trap. The remaining procedure depends on whether the final magnetic field $B$, at which the gas is imaged, is above or below 840 G.

If the final $B\geqslant840$ G, the evaporative cooling is performed at 840 G. This allows to achieve the lowest temperatures and ensures thermal equilibrium at higher final fields, where collisions are less frequent. At 1 s after the MOT is off, forced evaporation commences: The trap depth lowers by a factor of 60 as $e^{-t/(3\text{ s})}$. In the single-beam dipole trap at depth below 40 $\mu$K, the axial confinement is dominated by the small curvature of the magnetic field, which gives trapping frequency 12.6 Hz at 840 G and compresses the cloud in the axial direction. At 14 s after the MOT is off, the standing wave is reestablished. 1 s later, forced evaporation continues: Trap lowers by a factor of 30--55.8 exponentially for 11 s with the time constant 2.73--3.23 s. Immediately after, the potential is increased by a factor of 9--79 following the $e^{t/(0.2\text{ s})}$ law. At 26.4--26.7 s after the MOT is off, the magnetic field is set to its final value $B$. At 27.3--27.4 s, the standing-wave dipole trap is turned off abruptly in $\simeq1$ $\mu$s, the gas expands for the time $t_{\text{exp}}=150$ $\mu$s,
and then the density distribution is photographed.

For experiments with lower final fields, $B<840$ G, the evaporation is done at the same field as the imaging is. At 1 s after the MOT is off, the forced evaporation begins: The trap depth decreases as $e^{-t/(3\text{ s})}$ by a factor of 80. At 14.8 s after the MOT is off, the standing wave is reestablished; at 15.4 s, the magnetic field is switched from 840 G to the final value 730--830 G; and at 15.6 s, the second stage of forced evaporation begins: The potential exponentially decreases by a factor of 2.5 with the time constant 1.5 s; then between 17 and 23.9 s the potential decreases by a factor of 13.5--20 with the time constant 1.76--1.96 s. Afterwards, the potential is raised by a factor of 2.5--17 over 3 s exponentially with the time constant 0.11--0.31 s.
We checked, using an ideal-Fermi-gas model, that this rise does not excite collective modes. After the rise, the potential is held still for 100 ms in the majority of the experiments, while in a few cases the hold time is 4--18 ms.
Finally, the lattice turns off, the gas expands for $t_{\text{exp}}=150$--$250$ $\mu$s,
and the density distribution is measured.

Frequencies $\omega_x$, $\omega_y$, and $\omega_z$ are measured by observing parametric resonances at $B=528$ G, where the gas is noninteracting. The anharmonicity in the $x$ and $y$ directions is accounted for by comparing excitation spectrum to exact calculations of an ideal-gas dynamics, while $\omega_z$ is found by comparing resonant frequency to the splitting between the Bloch bands. The most shallow traps are used in the Bose regime where the chemical potential is small and, therefore, the anharmonic corrections are least important.

\section{Linear and planar density distributions}\label{sec:ImagingDensityProfiles}

\begin{figure}[htb!]
\begin{center}
\includegraphics[width=3.2in]{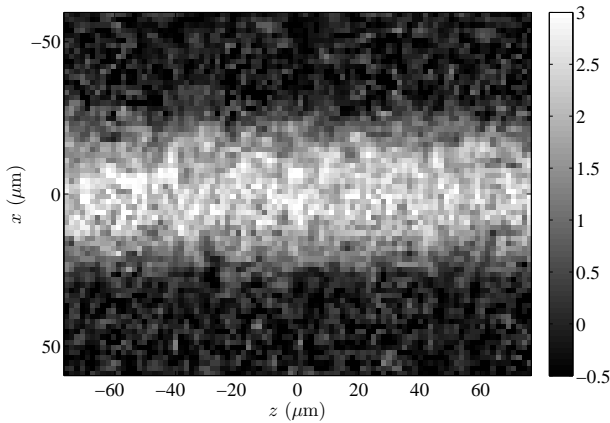}
\end{center}
\caption{An example of the column density distribution $n_{2\,\text{col}}(x,z)$ measured after the release and short expansion, which leaves profile along $x$ nearly unchanged. Tones of gray reflect the density value in atoms per $\mu$m$^2$ per spin as shown on the right. Fig.~\ref{fig:DensityProfiles} has been obtained from this image.}
\label{fig:ColDensity}
\end{figure}

The absorption imaging technique is employed: The gas is irradiated by a 3.7 $\mu$s pulse of a uniform laser beam resonant to a cycling two-level transition for one of the two spin states at the wavelength of $\lambda=671$ nm. The intensity is 1.3--1.7 mW/cm$^2=(0.50$--$0.65)I_{\text{sat}}$, where  $I_{\text{sat}}=2.54$ mW/cm$^2$ is the saturation intensity. The shadow, which the atoms make in the imaging beam, is projected and recorded on a CCD camera. From the shadow, we reconstruct the column density distribution $n_{2\,\text{col}}(x,z)$~\cite{FermionsInLattice,Fermi2D}. An example of such distribution is shown in Fig.~\ref{fig:ColDensity}.
The linear density profiles, as in Fig.~\ref{fig:DensityProfiles}(a), are obtained by averaging over the central $M=15$--$30$ nearly identical clouds, i. e. over 80--160 $\mu$m long region: $n_1(x)=(1/M)\int n_{2\,\text{col}}(x,z)\,dz$. The resolution, defined as the radius of the point spread function, is 1.9--2.9 $\mu$m depending on the $y$ size of the cloud. One pixel corresponds to a $1.7\times1.7$ $\mu$m$^2$ area at the atom location.

In all measurements, prior to imaging, the potential is turned off and the gas expands for the time $t_{\text{exp}}$ ranging from $2.4/\omega_z$ to $12.9/\omega_z$ and typically $\simeq6/\omega_z$. Despite the expansion, the imaging is nearly \textit{in situ} for the transverse direction, because $t_{\text{exp}}\ll1/\omega_x$. During $t_{\text{exp}}$, the transverse density distribution expands in a nearly selfsimilar way. For $t_{\text{exp}}=6/\omega_z$, various models predict a 0.2--0.4\% transverse expansion, with the lowest and highest values obtained within the BEC hydrodynamics~\cite{StringariHydro2002}
and ballistic model, respectively. In the data analysis, we account for the respective decrease of the central density. On the contrary, the expansion along $z$ is significant: For $t_{\text{exp}}=6/\omega_z$, each cloud expands along $z$ by a factor of 6--8 as calculated within the ballistic and unitary hydrodynamic~\cite{StringariHydro2002}
model, respectively. The modulation along $z$ nearly disappears, as in Fig.~\ref{fig:ColDensity}. This removes a potential systematic: If the clouds were too thin in the $z$ direction, the image would not fit into our NA=0.29 objective, reducing the apparent atom number~\cite{Fermi2D}. The rms cloud half-width along $z$ may be as large as 4 $\mu$m. Such significant cloud overlap does not compromise measured $n_1(x)$ distributions, because the 15--30 central clouds, which are used for analysis, and the clouds adjacent to them have nearly identical transverse profiles. Also the cloud overlap may be reducing the collective effects in light absorption, which are discussed below.

At the time of imaging, at the center of the joint cloud that appears from overlap of the initial isolated clouds, typical distance between same-spin atoms is $L=1.6$--$2.9\text{ $\mu$m}=(2.4$--$4.3)\lambda$. The optical density is in the range OD=0.25--1.1. Despite $L>\lambda$, we clearly observe collective effects in light absorption: As the gas keeps expanding for times $>1/\omega_x$, the apparent number of atoms $N_{\text{app}}$ keeps increasing saturating at interparticle distance in the densest part of the cloud $L\simeq13\,\lambda$ and $\text{OD}\simeq0.1$. It is not clear whether the collective absorption effect is local or depends on the cloud geometry. Assuming locality, we set up correction of $n_3$, the local density at each point of imaged gas,
\begin{equation}
\frac{n_{3\,\text{app}}}{n_3}=1-(0.34\pm0.02)(n_3\lambda^3)^{0.46\pm0.05},
\label{eq:LocalCorrection}
\end{equation}
where $n_{3\,\text{app}}$ is the apparent local density. As the alternative approach, we assume that collective absorption is nonlocal, and also that the long direction of the cloud is not a parameter of correction. In this case, the only parameter is the apparent optical density at the center, OD. Than the correction to the atom number takes form
\begin{equation}
\frac{N_{\text{app}}}N=1-(0.058\pm0.005)\text{OD}^{0.77\pm0.08}.
\label{eq:NonLocalCorrection}
\end{equation}
Formula~(\ref{eq:LocalCorrection}) assumes stronger corrections to the cloud center than to the edges, while (\ref{eq:NonLocalCorrection}) corrects density distribution by a uniform factor. Each formula reproduces correction to the atom number $N$ nearly equally well. Therefore, there is no clear choice, whether local or nonlocal formula has to be applied. We apply each formula with 50\% weight to correcting $n_2$ and $P_2/P_{2\,\text{ideal}}$. The half-difference between these two corrections is taken as a systematic error. Typical correction to $P_2/P_{2\,\text{ideal}}$ is $-6\%$. The following example may serve for estimating the difference of these two corrections: In the Fermi region, when solely local correction~(\ref{eq:LocalCorrection}) is applied, $P_2/P_{2\,\text{ideal}}$ is 8.5\% above the Fermi-liquid model~\cite{Bloom1975}, while for nonlocal correction~(\ref{eq:NonLocalCorrection}) the normalized pressure is 12.3\% above the model.

On the Bose side of the Feshbach resonance ($730\text{ G}<B<832$ G), formation of the Feshbach molecules does not reduce the visible number of atoms in comparison to that on the Fermi side. This is consistent with
Ref.~\cite{HuletSpectr},
where $\sim10^{-3}$ closed-channel fraction was found at $B=730$ G.

The planar density $n_2(\vec\rho)$, which is needed for calculating $P_{2\,\text{ideal}}$, may be obtained from $n_1(x)$ by means of the inverse Abel transform. Prior to the transform, noise in $n_1(x)$ with spatial period smaller than $0.4$ of the apparent Thomas-Fermi radius is filtered out. Such filtering leaves the apparent temperatures nearly unchanged. Since the original Abel transform is for cylindrical objects while the clouds are elliptic ($\omega_y/\omega_x=1.50$), we stretch the $y$ coordinate $y\rightarrow\tilde{y}\equiv y\,\omega_y/\omega_x$. Then the inverse Abel transform is applied as
\begin{equation}
n_2(\tilde{\rho})=-\frac{\omega_y/\omega_x}\pi\int_{\tilde{\rho}}^\infty\frac{dn_1}{dx}\frac{dx}{\sqrt{x^2-\tilde{\rho}^2}},
\label{eq:InvAbel}
\end{equation}
where $\tilde{\rho}\equiv\sqrt{x^2+\tilde{y}^2}$. Example of the resulting distribution $n_2(\tilde{\rho})$ is shown in Fig.~\ref{fig:DensityProfiles}(b) derived from the data of Fig.~\ref{fig:DensityProfiles}(a). To find the density at the origin we fit the data $n_2(\tilde{\rho})$ by a parabola within $\simeq75\%$ of the apparent Thomas-Fermi radius.

\begin{figure}[htb!]
\begin{center}
\includegraphics[width=2.5in]{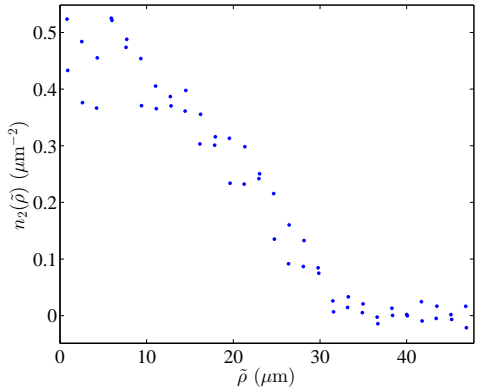}
\end{center}
\caption{Surface density $n_2(\tilde{\rho})$ derived from $n_1(x)$ of Fig.~\ref{fig:DensityProfiles}(a) without noise filtering. Compare to Fig.~\ref{fig:DensityProfiles}(b).}
\label{fig:PlanarDensityNoFiltering}
\end{figure}

To understand the reason for and consequences of applying the high-frequency noise filter to $n_1(x)$ consider $n_2(\tilde{\rho})$ of Fig.~\ref{fig:PlanarDensityNoFiltering} obtained from the same $n_1(x)$ [Fig.~\ref{fig:DensityProfiles}(a)] without noise filtering.
The noise is largest near the origin $\tilde{\rho}=0$, where the integrand in~(\ref{eq:InvAbel}) diverges. Fig.~\ref{fig:PlanarDensityNoFiltering} is among the least noisy. In different experiments, the rms noise on $n_1(x)$ is typically $3\%$ of $n_1(0)$ and nearly always within 1.5--4\%.
If high-frequency noise in $n_1(x)$ were left unfiltered, the rms noise in $n_2(\tilde{\rho})$ would be 10--30\% of the central density $n_2$, when calculated for $\tilde{\rho}$ within $75\%$ of the apparent Thomas-Fermi radius.
Modeling shows that such noise in $n_2(\tilde{\rho})$ would correspond to quite high 5--16\% standard error in $n_2$, when $n_2$ is determined from the parabolic fit. Prefiltering of $n_1(x)$ reduces the fit error significantly. For example, modeling predicts reduction of the fit error in $n_2$ down to 1.2--4.4\%.

\section{Error analysis}\label{sec:ErrAnalysis}

For the purpose of error analysis, we write the normalized pressure without the anharmonic correction:
\begin{equation}
\frac{P_2}{P_{2\,\text{ideal}}}\simeq\frac m{2\pi^2\hbar^2}\frac{\omega_x^3}{\omega_y}\frac N{\tilde{n}_2^2},
\end{equation}
where we have introduced quantity $\tilde{n}_2=n_2\,\omega_x/\omega_y$ whose uncertainty is independent of the errors in the frequencies as one may see in Eq.~(\ref{eq:InvAbel}). Let $\delta[...]$ designate the relative uncertainty of quantity in square parentheses. The uncertainty of the normalized pressure may be expressed via the uncertainty of the two uncorrelated quantities:
\begin{equation}
\delta\!\left[\frac{P_2}{P_{2\,\text{ideal}}}\right]= \sqrt{\left(\delta\!\left[\frac{\omega_x^3}{\omega_y}\right]\right)^2+\left(\delta\!\left[\frac N{\tilde{n}_2^2}\right]\right)^2}.
\label{eq:deltaP}
\end{equation}

The frequencies are measured in experiments that are separate from measuring the density profiles $n_1(x)$. The frequency errors influence equally each measurement of the normalized pressure. The longitudinal frequency $\omega_z$ is known within $\pm2\%$, while $\omega_x$ and $\omega_y$ are known within $\pm0.5\%$. In the frequency uncertainties, we account for slow drifts between successive measurements. In Eq.~(\ref{eq:deltaP}), the frequency-related uncertainty is $\delta\!\left[\frac{\omega_x^3}{\omega_y}\right]=1.6\%$.

The pressure and density have been measured in 271 repetitions of experiment for $a_2\sqrt{n_2}=0.0005$--$64$. The value $a_2$ is varied by changing the magnetic field and trap depth. Among the measurements at the same $B$ and $l_z$, outcomes with close $a_2\sqrt{n_2}$ values are grouped together. For each group the mean is reported on Figs.~\ref{fig:PvskFa2} and \ref{fig:BoseLinear} and in the Table. Each group contains $K=2$--18 measurements (7 on average, see Table).
For the $i$th measurement we deduce $N_i$ and $\tilde{n}_{2i}$, from which the mean and the statistical uncertainty are calculated:
\begin{equation}
\frac N{\tilde{n}_2^2}=\frac1K\sum_{i=1}^K\frac{N_i}{\tilde{n}_{2i}^2},
\end{equation}
\begin{equation}
\delta\!\left[\frac N{\tilde{n}_2^2}\right]_{\text{stat}}= \sqrt{\frac1{K(K-1)}\sum_{i=1}^K\left(\frac{N_i}{\tilde{n}_{2i}^2}\frac{\tilde{n}_2^2}N-1\right)^2}.
\end{equation}
The total uncertainty $\delta\!\left[\frac N{\tilde{n}_2^2}\right]$ is the combination of systematic and statistical errors:
\begin{equation}
\delta\!\left[\frac N{\tilde{n}_2^2}\right]=\sqrt{\left(\delta\!\left[\frac N{\tilde{n}_2^2}\right]_{\text{sys}}\right)^2+\left(\delta\!\left[\frac N{\tilde{n}_2^2}\right]_{\text{stat}}\right)^2}.
\end{equation}
The statistical error appears from the fit error in determination of $\tilde{n}_2$, from the imaging-laser-frequency fluctuations as well as for undetermined reasons. These error sources are all together reflected in the data scatter in each group of $K$ measurements. The value $\delta\!\left[\frac N{\tilde{n}_2^2}\right]_{\text{stat}}$ lies in the range 2--8\% and is 3.5\% on average.

The systematic error appears from the uncertainties in corrections to systematic effects.
The atom-counting uncertainties $\delta[N]_{\text{sys}}$ are equally applied to $N$ and $\tilde{n}_2$ and therefore partially cancel in $N/\tilde{n}_2^2$. One may write
\begin{equation}
\delta\!\left[\frac N{\tilde{n}_2^2}\right]_{\text{sys}}\!=\sqrt{\left(\delta\!\left[\frac N{\tilde{n}_2^2}\right]_{\text{abs}}\right)^2+(\delta[N]_{\text{sys}})^2+(2\,\delta[\tilde{n}_2]_{\text{filt}})^2},
\end{equation}
where $\delta\!\left[\frac N{\tilde{n}_2^2}\right]_{\text{abs}}$ is the systematic error arising from the difference between the two protocols for correcting the collective absorption effect [Eqs.~(\ref{eq:LocalCorrection}) and (\ref{eq:NonLocalCorrection})]; $\delta[\tilde{n}_2]_{\text{filt}}=0.8\%$ is the systematic-uncertainty part unrelated to atom counting and arising from the high-frequency filter, which is applied to $n_1(x)$ prior to the inverse Abel transform. The quadratically combined dominant sources for $\delta[N]_{\text{sys}}$ are:
(i)~1.5\%~-- the uncertainty of incorrect imaging-beam polarization;
(ii)~1\%~-- the uncertainty of dark and stray-light counts on the CCD;
(iii)~1\%~-- the fluctuations and uncertainty of the imaging-laser frequency, which reduce the apparent atom number by $2\pm1\%$;
and (iv)~$0.9\%$~-- the uncertainty of losses on the imaging optics.
As a result, $\delta[N]_{\text{sys}}=2.5\%$. The total systematic error $\delta\!\left[\frac N{\tilde{n}_2^2}\right]_{\text{sys}}$ is in the range 3.2--4.1\% and 3.5\% on average.

The combined uncertainty $\delta\!\left[\frac{P_2}{P_{2\,\text{ideal}}}\right]$ is in the range 3.7--9\% and 5.3\% on average.
Fig. 4 is representative with regards to checking the correctness of the error bars. Fig. 4 shows that for 62\% of data, the fitting curve falls into the $\pm1\sigma$ interval, which is near the 68\% statistically required for comparison between correct experiment and correct theory. This supports the correctness of our choice of the error bars.

The uncertainty in $a_2\sqrt{n_2}$ is the quadratic combination of its statistical standard error, systematic uncertainty of $n_2$, and systematic uncertainty of $a_2$. The latter arises from $\delta[B]_{\text{sys}}\simeq0.1\%$ and from $\delta[\mu]_{\text{sys}}$. The source of $\delta[\mu]_{\text{sys}}$ is our assumption $\mu\propto n_2$ used in calculating $\mu$ and $a_2$. The more general assumption is $\mu\propto n_2^\gamma$, which gives $\mu=\frac{\gamma+1}{2\gamma}\sqrt{\frac{P_2}{P_{2\,\text{ideal}}}}E_F$ for a 2D gas in a parabolic potential. This formula lets one calculate $\delta[\mu]_{\text{sys}}$ provided the departure of $\gamma$ from 1 is known. From the curvature of $n_2(\tilde{\rho})$ [Fig.~\ref{fig:DensityProfiles}(b)], we measure at the cloud center
\begin{equation}
\gamma=\frac{dP_2}{dn_2}\frac{n_2}{P_2}-1=\frac{m\omega_x^2n_2^2}{n_2''P_2}-1
\end{equation}
and find that the $\gamma$ values are scattered around values slightly above 1 depending on the interaction. This makes us take $\gamma=1^{+0.07}_{-0.03}$, $1^{+0.1}_{-0}$, and $1^{+0.14}_{-0}$ in the Bose, strongly interacting, and Fermi regime respectively. This induces asymmetric systematic uncertainties on $\mu$ and consequently on $a_2$. Without $\delta[\mu]_{\text{sys}}$, the typical total standard error on $a_2$ is $\pm3.6\%$. Taking $\delta[\mu]_{\text{sys}}$ into account increases the typical error by a small value, up to $^{+4.6}_{-3.6}\%$.

\section{Fitting data of Fig.~\ref{fig:PvskFa2} by a smooth curve $p_{\text{fit}}(a_2\sqrt{n_2})$}\label{sec:FittingByCurve}

\begin{figure}[htb!]
\begin{center}
\includegraphics[width=3.2in]{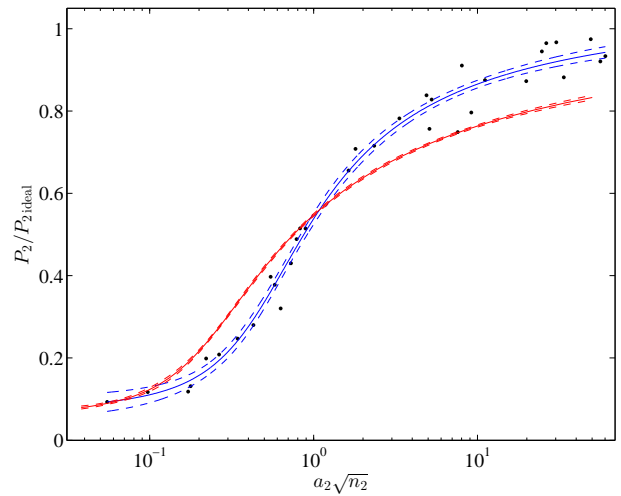}
\end{center}
\caption{Normalized local pressure vs coupling parameter. Black dots: The data. The fit to the data is shown by the solid blue curve with the dashed curves displaying the fit uncertainty. The fit to the Monte Carlo simulation~\cite{Giorgini2D2011} is shown by the solid red curve with the dashed curves displaying the fit uncertainty.}
\label{fig:PvskFa2Fit}
\end{figure}

The data of Fig.~\ref{fig:PvskFa2} is fitted in the range $a_2\sqrt{n_2}=0.055$--60 by function
\begin{equation}
p_{\text{fit}}(a_2\sqrt{n_2})= A_1\left[\frac\pi2+\text{arctg}\!\left[A_2\ln(a_2\sqrt{n_2})+A_3\right]\right]^2+A_4,
\label{eq:FitFunction}
\end{equation}
where $A_i$ are the fit parameters, which converge to values $A_1=0.104$, $A_2=0.85$, $A_3=0.62$, and $A_4=0.07$. The fit is shown in Fig.~\ref{fig:PvskFa2Fit}.

This fit is used in producing Fig.~\ref{fig:muDoesntMatter}, where the ratio of the measured pressure to the fit is shown. Also the fit is useful for demonstrating the deviation between the data and the pure 2D Monte Carlo simulation~\cite{Giorgini2D2011} in the strongly interacting region. In particular, in Figs.~\ref{fig:PvskFa2} and \ref{fig:PvskFa2Fit}, Monte Carlo simulation~\cite{Giorgini2D2011} is approximated by function~(\ref{eq:FitFunction}) with other $A_i$ values. On Fig.~\ref{fig:PvskFa2Fit}, in the region of strong interactions, one may compare the fit to the data and the fit to the Monte Carlo simulation: The slope of the fit to the data is larger than the slope of the fit to the simulation by more than the fit uncertainties.

\section{Data representation for different definitions of $a_2$}\label{sec:DataReps}

\begin{figure}[htb!]
\begin{center}
\includegraphics[width=3.2in]{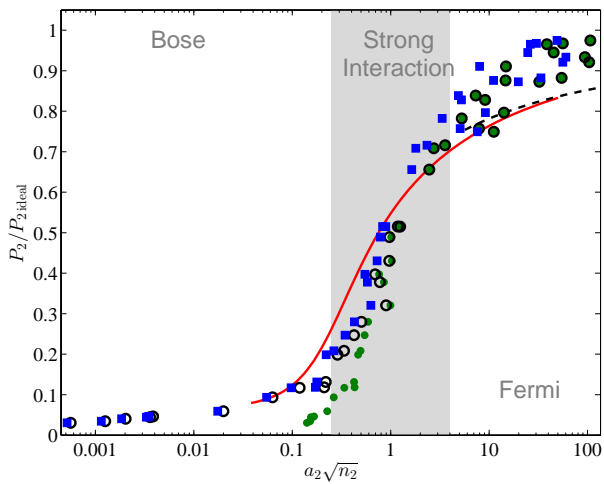}
\end{center}
\caption{Normalized local pressure vs coupling parameter plotted for different definitions of $a_2$. Blue squares: $a_2$ defined as in the main text, i.~e. same representation as in Figs.~\ref{fig:PvskFa2} and \ref{fig:BoseLinear}. Black circles: $a_2$ defined by Eq.~(\ref{eq:a2}) as in Ref.~\cite{ValeCloudSize2011}. Green dots: $a_2$ defined from the bound-state energy and Eq.~(\ref{eq:Eb}) similar to Refs.~\cite{Kohl2DPseudogap2011,Kohl2DFermiLiquid2012}. Solid red curve: The smooth approximation to the Monte Carlo simulation for a purely 2D system~\cite{Giorgini2D2011}. Dashed curve: The Fermi-liquid theory for $T=0$~\cite{Bloom1975}.}
\label{fig:PvskFa2Alternatives}
\end{figure}

We find $a_2$ by matching the two-body scattering problem in the 2D space and in the tight harmonic potential along $z$. The value of $a_2$ is calculated by equating the scattering amplitudes $f_{\text{2D}}(q,a_2)=f_{\text{Q2D}}(q,a,l_z)$ [Eqs.~(\ref{eq:f2}) and (\ref{eq:f2ho})], where $w(\xi)$ is defined as in~\cite{Shlyapnikov2DScattering2001}
\begin{equation}
w(\xi)\!\equiv\!\lim_{J\rightarrow\infty}\!\left[\sqrt{\frac{4J}\pi}\ln\frac J{e^2}- \sum_{j=0}^J\frac{(2j-1)!!}{(2j)!!}\ln(j-\xi-i0)\right]\!\!.
\label{eq:wOfxi}
\end{equation}
In the limit $\xi\ll1$ ($q^2l_z^2\ll1$), one may approximate $w(\xi)\simeq-\ln(2\pi \xi/0.905)+i\pi$, which gives the energy independent expression, used in Ref.~\cite{ValeCloudSize2011}:
\begin{equation}
a_2\simeq2.96\,l_z\,e^{-\sqrt\pi l_z/a}.
\label{eq:a2}
\end{equation}
In Fig.~\ref{fig:PvskFa2Alternatives}, we replot our pressure data using the definition~(\ref{eq:a2}) for $a_2$ through out all regimes, as in Ref.~\cite{ValeCloudSize2011}: In this case, the pressure is systematically below the Monte Carlo simulation~\cite{Giorgini2D2011} in the most of the strong-interaction region.

Alternatively, $a_2$ is found from binding energy $E_{\text{3D bound}}$ of the 3D dimer molecule in the potential $(2m)\omega_z^2z^2/2$. The energy comes from equation~\cite{BlochLowDReview2008}
\begin{equation}
\frac{l_z}{a}=\int_0^\infty\frac{du}{\sqrt{8\pi u^3}}\left(1- \frac{e^{-u|E_{\text{3D bound}}|/\hbar\omega_z}}{\sqrt{(1-e^{-2u})/2u}}\right).
\label{eq:Eb}
\end{equation}
Then $a_2$ is found by equating $E_{\text{3D bound}}$ to the binding energy in a 2D potential: $E_{\text{3D bound}}=-4\hbar^2/me^{2\gamma_E}a_2^2$.

For comparison, in Fig.~\ref{fig:PvskFa2Alternatives} we have also drawn our data using the $a_2$ calibration based on Eq.~(\ref{eq:Eb}). This option of defining $a_2$ was used in Refs.~\cite{Kohl2DPseudogap2011,Kohl2DFermiLiquid2012} up to a constant factor $\simeq1$. In the Fermi regime, the approaches based on Eqs.~(\ref{eq:a2}) and (\ref{eq:Eb}) give nearly identical $a_2$ calibration, which differ from the approach adopted by us, because the latter accounts for finite kinetic energy of colliding particles. In the Bose regime, our scattering-amplitude-based approach and the binding-energy approach give orders-of-magnitude difference in $a_2$ as seen in Fig.~\ref{fig:PvskFa2Alternatives}.



\begin{table*}{\footnotesize
\begin{tabular}[t]{|r@{}@{}l|c|r@{}@{}l|c|c|c|c|c|c|p{12.8em}|p{11.3em}|}
\hline
\multicolumn{2}{|c|}{$a_{2}\sqrt{n_2}$} & \parbox{3.0em}{$a_2$ (Bohr)} & \multicolumn{2}{|c|}{$P_2/P_{2\text{ ideal}}$} & \parbox{3.3em}{$B$ (Gauss)} & $s$ & \parbox{3.0em}{$a$ (Bohr)} & $N$ & $T/\varepsilon_F$ & \parbox{2.8em}{\# of repetitions} & $N$ in each repetition & $T/\varepsilon_F$ in each repetition\\
\hline
 (5.1$\pm$&.4)$10^{-4}$ & 10 & 0.031$\pm$ & 0.002 & 730 & 15 & 2540 & 710 &
0.036 & 3 & 700,\hspace{0.1em}720,\hspace{0.1em}690 &
.04,\hspace{0.1em}.03,\hspace{0.1em}.04 \\
 (11.5$\pm$&.8)$10^{-4}$ & 21 &
0.035$\pm$ & 0.002 & 730 & 23.3 & 2540 & 690 & 0.04 & 5 &
640,\hspace{0.1em}730,\hspace{0.1em}670,\hspace{0.1em}690,\hspace{0.1em}720 &
.04,\hspace{0.1em}.04,\hspace{0.1em}.03,\hspace{0.1em}.02,\hspace{0.1em}.04
\\
 (1.8$\pm$&.1)$10^{-3}$ &     34 &  0.041$\pm$ &  0.002 &  730 &   31.8 &   2540 &  660 & 0.04 & 3 & 640,\hspace{0.1em}720,\hspace{0.1em}630 & .04,\hspace{0.1em}.05,\hspace{0.1em}.04 \\
 (3.3$\pm$&.2)$10^{-3}$ &     60 &  0.044$\pm$ &  0.002 &  730 &     49 &   2540 &  470 & 0.02 & 2 & 430,\hspace{0.1em}510 & .03,\hspace{0.1em}.00 \\
 (3.5$\pm$&.2)$10^{-3}$ &     59 &  0.046$\pm$ &  0.003 &  730 &     49 &   2540 &  640 & 0.044 & 7 & 680,\hspace{0.1em}690,\hspace{0.1em}580,\hspace{0.1em}630,\hspace{0.1em}650,\hspace{0.1em}610,\hspace{0.1em}670 & .04,\hspace{0.1em}.05,\hspace{0.1em}.05,\hspace{0.1em}.03,\hspace{0.1em}.05,\hspace{0.1em}.04,\hspace{0.1em}.04 \\
 (17.5$\pm$&.7)$10^{-3}$ &    299 &  0.059$\pm$ &  0.003 &  750 &     49 &   3520 &  800 & 0.05 & 12 & 790,\hspace{0.1em}800,\hspace{0.1em}780,\hspace{0.1em}850,\hspace{0.1em}830,\hspace{0.1em}810,\hspace{0.1em}790,\hspace{0.1em}790,\hspace{0.1em}820,\hspace{0.1em}770,\hspace{0.1em}810,\hspace{0.1em}730 & .05,\hspace{0.1em}.04,\hspace{0.1em}.06,\hspace{0.1em}.06,\hspace{0.1em}.05,\hspace{0.1em}.05,\hspace{0.1em}.04,\hspace{0.1em}.05,\hspace{0.1em}.04,\hspace{0.1em}.06,\hspace{0.1em}.06,\hspace{0.1em}.04 \\
 (5.5$\pm$&.2)$10^{-2}$ &   1150 &  0.093$\pm$ &  0.004 &  770 &     49 &   5100 &  560 & 0.06 & 9 & 580,\hspace{0.1em}600,\hspace{0.1em}540,\hspace{0.1em}580,\hspace{0.1em}580,\hspace{0.1em}640,\hspace{0.1em}490,\hspace{0.1em}560,\hspace{0.1em}490 & .06,\hspace{0.1em}.07,\hspace{0.1em}.06,\hspace{0.1em}.05,\hspace{0.1em}.06,\hspace{0.1em}.06,\hspace{0.1em}.07,\hspace{0.1em}.06,\hspace{0.1em}.08 \\
  (9.7$\pm$&.3)$10^{-2}$ &   1940 &  0.117$\pm$ &  0.006 &  780 &     49 &   6400 &  870 & 0.068 & 7 & 820,\hspace{0.1em}990,\hspace{0.1em}850,\hspace{0.1em}780,\hspace{0.1em}860,\hspace{0.1em}880,\hspace{0.1em}920 & .07,\hspace{0.1em}.07,\hspace{0.1em}.07,\hspace{0.1em}.04,\hspace{0.1em}.07,\hspace{0.1em}.07,\hspace{0.1em}.08 \\
  (17.2$\pm$&.7)$10^{-2}$ &   4600 &   0.12$\pm$ &   0.01 &  800 &     15 &  11300 &  910 & 0.07 & 3 & 940,\hspace{0.1em}960,\hspace{0.1em}840 & .06,\hspace{0.1em}.07,\hspace{0.1em}.08 \\
  0.178 & $_{- 0.003}^{+ 0.004}$  &   4700 &  0.132$\pm$ &  0.005 &  800 &   18.4 &  11300 &  830 & 0.07 & 13 & 700,\hspace{0.1em}850,\hspace{0.1em}860,\hspace{0.1em}850,\hspace{0.1em}830,\hspace{0.1em}840,\hspace{0.1em}840,\hspace{0.1em}830,\hspace{0.1em}820,\hspace{0.1em}770,\hspace{0.1em}740,\hspace{0.1em}890,\hspace{0.1em}910 & .08,\hspace{0.1em}.04,\hspace{0.1em}.08,\hspace{0.1em}.07,\hspace{0.1em}.08,\hspace{0.1em}.07,\hspace{0.1em}0,\hspace{0.1em}.09,\hspace{0.1em}.06,\hspace{0.1em}.06,\hspace{0.1em}.08,\hspace{0.1em}.06,\hspace{0.1em}.09 \\
  0.221 & $_{- 0.006}^{+ 0.007}$  &   5100 &    0.2$\pm$ &   0.01 &  800 &     49 &  11300 &  850 & 0.097 & 6 & 650,\hspace{0.1em}920,\hspace{0.1em}820,\hspace{0.1em}900,\hspace{0.1em}800,\hspace{0.1em}1000 & .11,\hspace{0.1em}.1,\hspace{0.1em}.07,\hspace{0.1em}.11,\hspace{0.1em}.09,\hspace{0.1em}.1 \\
  0.265 & $_{- 0.006}^{+ 0.007}$  &   5400 &   0.21$\pm$ &   0.01 &  800 &    102 &  11300 &  660 & -- & 6 & 780,\hspace{0.1em}660,\hspace{0.1em}600,\hspace{0.1em}620,\hspace{0.1em}650,\hspace{0.1em}670 & -- \\
   0.34&$\pm$0.01  &   8200 &  0.247$\pm$ &  0.013 &  810 &    101 &  17100 &  450 & -- & 7 & 350,\hspace{0.1em}390,\hspace{0.1em}510,\hspace{0.1em}460,\hspace{0.1em}450,\hspace{0.1em}460,\hspace{0.1em}490 & -- \\
   0.43&$\pm$0.02  &  12100 &   0.28$\pm$ &   0.02 &  820 &    101 &  32500 &  254 & -- & 5 & 180,\hspace{0.1em}190,\hspace{0.1em}310,\hspace{0.1em}320,\hspace{0.1em}260 & -- \\
   0.55&$\pm$0.02  &  15400 &    0.4$\pm$ &   0.03 &  830 &    101 & 189000 &  350 & -- & 3 & 320,\hspace{0.1em}390,\hspace{0.1em}340 & -- \\
   0.58 & $_{- 0.015}^{+  0.02}$  &  14600 &   0.38$\pm$ &   0.02 &  830 &    101 & 189000 &  530 & -- & 6 & 610,\hspace{0.1em}490,\hspace{0.1em}500,\hspace{0.1em}540,\hspace{0.1em}500,\hspace{0.1em}530 & -- \\
   0.63 & $_{- 0.015}^{+  0.02}$  &  16200 &  0.321$\pm$ &  0.014 &  830 &     49 & 189000 &  880 & -- & 5 & 880,\hspace{0.1em}840,\hspace{0.1em}890,\hspace{0.1em}900,\hspace{0.1em}870 & -- \\
   0.73 & $_{- 0.015}^{+  0.02}$  &  19800 &   0.43$\pm$ &   0.02 &  840 &    101 & -55000 &  450 & -- & 4 & 460,\hspace{0.1em}440,\hspace{0.1em}470,\hspace{0.1em}420 & -- \\
   0.79&$\pm$0.04  &  21600 &   0.49$\pm$ &   0.04 &  850 &    210 & -24900 &  235 & -- & 9 & 170,\hspace{0.1em}200,\hspace{0.1em}190,\hspace{0.1em}180,\hspace{0.1em}220,\hspace{0.1em}295,\hspace{0.1em}289,\hspace{0.1em}294,\hspace{0.1em}260 & -- \\
   0.83&$\pm$0.03  &  18600 &   0.52$\pm$ &   0.03 &  850 &    211 & -24900 &  540 & -- & 5 & 540,\hspace{0.1em}540,\hspace{0.1em}550,\hspace{0.1em}520,\hspace{0.1em}550 & -- \\
   0.89 & $_{-  0.02}^{+  0.03}$  &  24800 &   0.51$\pm$ &   0.02 &  850 &    102 & -24900 &  490 & -- & 18 & 510,\hspace{0.1em}510,\hspace{0.1em}510,\hspace{0.1em}490,\hspace{0.1em}510,\hspace{0.1em}500,\hspace{0.1em}510,\hspace{0.1em}520,\hspace{0.1em}520,\hspace{0.1em}500,\hspace{0.1em}470,\hspace{0.1em}500,\hspace{0.1em}540,\hspace{0.1em}450,\hspace{0.1em}460,\hspace{0.1em}450,\hspace{0.1em}430,\hspace{0.1em}460 & -- \\
   1.63 & $_{-  0.05}^{+  0.06}$  &  47000 &   0.66$\pm$ &   0.03 &  880 &    101 & -10300 &  540 & -- & 10 & 550,\hspace{0.1em}530,\hspace{0.1em}550,\hspace{0.1em}540,\hspace{0.1em}560,\hspace{0.1em}500,\hspace{0.1em}550,\hspace{0.1em}550,\hspace{0.1em}550,\hspace{0.1em}520 & -- \\
    1.8 & $_{-  0.07}^{+   0.1}$  &  44000 &   0.71$\pm$ &   0.04 &  900 &    211 &  -7700 &  510 & -- & 9 & 550,\hspace{0.1em}520,\hspace{0.1em}530,\hspace{0.1em}490,\hspace{0.1em}530,\hspace{0.1em}520,\hspace{0.1em}500,\hspace{0.1em}490,\hspace{0.1em}490 & -- \\
    2.3 & $_{-   0.1}^{+  0.13}$  &  70000 &   0.72$\pm$ &   0.05 &  900 &    102 &  -7700 &  520 & -- & 5 & 520,\hspace{0.1em}520,\hspace{0.1em}570,\hspace{0.1em}540,\hspace{0.1em}440 & -- \\
    3.3 & $_{-  0.15}^{+   0.2}$  &  84000 &   0.78$\pm$ &   0.04 &  950 &    211 &  -5200 &  530 & -- & 4 & 540,\hspace{0.1em}580,\hspace{0.1em}520,\hspace{0.1em}500 & -- \\
    4.9&$\pm$0.2  & 161000 &   0.84$\pm$ &   0.04 &  950 &    101 &  -5200 &  400 & 0.08 & 7 & 430,\hspace{0.1em}430,\hspace{0.1em}390,\hspace{0.1em}410,\hspace{0.1em}380,\hspace{0.1em}420,\hspace{0.1em}370 & .13,\hspace{0.1em}.02,\hspace{0.1em}.05,\hspace{0.1em}.02,\hspace{0.1em}.13,\hspace{0.1em}.15,\hspace{0.1em}.03 \\
    5.1 & $_{-   0.2}^{+   0.3}$  & 154000 &   0.76$\pm$ &   0.03 &  950 &    102 &  -5200 &  520 & 0.08 & 3 & 500,\hspace{0.1em}530,\hspace{0.1em}520 & .05,\hspace{0.1em}.08,\hspace{0.1em}.12 \\
    5.2 & $_{-   0.2}^{+   0.4}$  & 127000 &   0.83$\pm$ &   0.03 & 1000 &    211 &  -4100 &  660 & 0.11 & 10 & 730,\hspace{0.1em}740,\hspace{0.1em}770,\hspace{0.1em}600,\hspace{0.1em}580,\hspace{0.1em}640,\hspace{0.1em}610,\hspace{0.1em}650,\hspace{0.1em}660,\hspace{0.1em}630 & .09,\hspace{0.1em}.09,\hspace{0.1em}.08,\hspace{0.1em}.13,\hspace{0.1em}.15,\hspace{0.1em}.13,\hspace{0.1em}.12,\hspace{0.1em}.12,\hspace{0.1em}.11,\hspace{0.1em}.09 \\
    7.6 & $_{-   0.3}^{+   0.4}$  & 240000 &   0.75$\pm$ &   0.03 &  980 &    101 &  -4500 &  430 & 0.07 & 2 & 420,\hspace{0.1em}430 & .03,\hspace{0.1em}.1 \\
      8 & $_{-   0.5}^{+   0.7}$  & 236000 &   0.91$\pm$ &   0.05 & 1000 &    102 &  -4100 &  710 & 0.1 & 3 & 680,\hspace{0.1em}700,\hspace{0.1em}740 & .12,\hspace{0.1em}.11,\hspace{0.1em}.07 \\
    9.2 & $_{-   0.4}^{+   0.5}$  & 286000 &    0.8$\pm$ &   0.03 & 1000 &    101 &  -4100 &  480 & 0.06 & 12 & 560,\hspace{0.1em}510,\hspace{0.1em}520,\hspace{0.1em}530,\hspace{0.1em}510,\hspace{0.1em}510,\hspace{0.1em}460,\hspace{0.1em}450,\hspace{0.1em}450,\hspace{0.1em}450,\hspace{0.1em}450,\hspace{0.1em}420 & .07,\hspace{0.1em}.03,\hspace{0.1em}.04,\hspace{0.1em}.07,\hspace{0.1em}.14,\hspace{0.1em}.03,\hspace{0.1em}.11,\hspace{0.1em}.04,\hspace{0.1em}.04,\hspace{0.1em}.02,\hspace{0.1em}.1,\hspace{0.1em}.07 \\
   11.1 & $_{-   0.6}^{+   0.7}$  & 284000 &   0.88$\pm$ &   0.06 & 1400 &    540 &  -2510 &  215 & 0.05 & 4 & 200,\hspace{0.1em}210,\hspace{0.1em}250,\hspace{0.1em}190 & .03,\hspace{0.1em}.02,\hspace{0.1em}.11,\hspace{0.1em}.03 \\
     20 & $_{-     1}^{+   1.3}$  & 620000 &   0.87$\pm$ &   0.04 & 1100 &    102 &  -3250 &  530 & 0.07 & 4 & 530,\hspace{0.1em}520,\hspace{0.1em}530,\hspace{0.1em}540 & .02,\hspace{0.1em}.09,\hspace{0.1em}.05,\hspace{0.1em}.13 \\
     25 & $_{-   1.4}^{+     2}$  & 620000 &   0.95$\pm$ &   0.04 & 1400 &    211 &  -2510 &  670 & 0.08 & 4 & 640,\hspace{0.1em}700,\hspace{0.1em}620,\hspace{0.1em}700 & .07,\hspace{0.1em}.05,\hspace{0.1em}.1,\hspace{0.1em}.08 \\
     26 & $_{-   1.4}^{+     2}$  & 790000 &   0.97$\pm$ &   0.06 & 1400 &    208 &  -2510 &  330 & 0.09 & 4 & 320,\hspace{0.1em}350,\hspace{0.1em}330,\hspace{0.1em}320 & .09,\hspace{0.1em}.05,\hspace{0.1em}.12,\hspace{0.1em}.1 \\
     30 & $_{-     2}^{+     3}$  & 920000 &   0.97$\pm$ &   0.04 & 1200 &    102 &  -2860 &  670 & 0.08 & 4 & 590,\hspace{0.1em}690,\hspace{0.1em}720,\hspace{0.1em}670 & .02,\hspace{0.1em}.1,\hspace{0.1em}.09,\hspace{0.1em}.09 \\
     34&$\pm$2  & 1050000 &   0.88$\pm$ &   0.05 & 1200 &    102 &  -2860 &  520 & 0.06 & 10 & 500,\hspace{0.1em}520,\hspace{0.1em}510,\hspace{0.1em}530,\hspace{0.1em}460,\hspace{0.1em}560,\hspace{0.1em}540,\hspace{0.1em}520,\hspace{0.1em}520,\hspace{0.1em}560 & .03,\hspace{0.1em}.05,\hspace{0.1em}.05,\hspace{0.1em}.08,\hspace{0.1em}.05,\hspace{0.1em}.15,\hspace{0.1em}.02,\hspace{0.1em}.03,\hspace{0.1em}.03,\hspace{0.1em}.1 \\
     49 & $_{-     4}^{+     6}$  & 1390000 &   0.97$\pm$ &   0.04 & 1400 &    105 &  -2510 &  880 & 0.07 & 8 & 1040,\hspace{0.1em}920,\hspace{0.1em}920,\hspace{0.1em}850,\hspace{0.1em}900,\hspace{0.1em}780,\hspace{0.1em}800,\hspace{0.1em}790 & .11,\hspace{0.1em}.07,\hspace{0.1em}.03,\hspace{0.1em}.07,\hspace{0.1em}.08,\hspace{0.1em}.05,\hspace{0.1em}.06,\hspace{0.1em}.06 \\
     56 & $_{-     3}^{+     5}$  & 1660000 &   0.92$\pm$ &   0.04 & 1400 &    103 &  -2510 &  700 & 0.06 & 16 & 720,\hspace{0.1em}760,\hspace{0.1em}670,\hspace{0.1em}740,\hspace{0.1em}700,\hspace{0.1em}750,\hspace{0.1em}730,\hspace{0.1em}720,\hspace{0.1em}690,\hspace{0.1em}610,\hspace{0.1em}670,\hspace{0.1em}630,\hspace{0.1em}680,\hspace{0.1em}650,\hspace{0.1em}730,\hspace{0.1em}670 & .03,\hspace{0.1em}.03,\hspace{0.1em}.08,\hspace{0.1em}.04,\hspace{0.1em}.12,\hspace{0.1em}.06,\hspace{0.1em}.04,\hspace{0.1em}.09,\hspace{0.1em}.07,\hspace{0.1em}.04,\hspace{0.1em}.06,\hspace{0.1em}.07,\hspace{0.1em}.06,\hspace{0.1em}.05,\hspace{0.1em}.02,\hspace{0.1em}.05 \\
     60 & $_{-     3}^{+     4}$  & 2020000 &   0.93$\pm$ &   0.04 & 1400 &    101 &  -2510 &  430 & 0.05 & 14 & 380,\hspace{0.1em}240,\hspace{0.1em}490,\hspace{0.1em}480,\hspace{0.1em}460,\hspace{0.1em}470,\hspace{0.1em}450,\hspace{0.1em}490,\hspace{0.1em}450,\hspace{0.1em}410,\hspace{0.1em}400,\hspace{0.1em}420,\hspace{0.1em}390,\hspace{0.1em}440 & .08,\hspace{0.1em}.11,\hspace{0.1em}.04,\hspace{0.1em}.03,\hspace{0.1em}.02,\hspace{0.1em}.04,\hspace{0.1em}.07,\hspace{0.1em}.03,\hspace{0.1em}.08,\hspace{0.1em}.06,\hspace{0.1em}.04,\hspace{0.1em}.02,\hspace{0.1em}.05,\hspace{0.1em}.05 \\
\hline
\end{tabular}}
\caption{Experimental results and parameters. The standard error is shown as indices when asymmetric. The first 5 points are not shown in Fig.~\ref{fig:PvskFa2} but are shown in Figs.~\ref{fig:BoseLinear} and \ref{fig:PvskFa2Alternatives}. The temperatures for the weakly interacting Fermi gas, $a_2\sqrt{n_2}=4.9$--$60$, are measured by fitting the Thomas-Fermi distribution of the ideal Fermi gas~\cite{Fermi2D} to the data, while the temperatures in the Bose regime, $a_2\sqrt{n_2}=0.00051$--$0.22$, are found from the bimodal fit.}
\end{table*}

\clearpage

\end{document}